\documentclass[journal,twoside,web]{ieeecolor}
\usepackage{generic}
\usepackage{cite}
\usepackage{amsmath,amssymb,amsfonts}
\usepackage{algorithmic}
\usepackage{graphicx}
\usepackage{textcomp}
\usepackage{multirow}
\usepackage{tabularx}
\usepackage{booktabs}
\usepackage{url}

\usepackage[utf8]{inputenc}
\DeclareUnicodeCharacter{22C5}{\ensuremath{{}^2}}
\DeclareUnicodeCharacter{2219}{\ensuremath{{}^2}}
\DeclareUnicodeCharacter{3BA}{\ensuremath{{}^2}}
\usepackage{etoolbox}
\makeatletter
\patchcmd{\@makecaption}
{\scshape}
{}
{}
{}
\makeatother

\def\BibTeX{{\rm B\kern-.05em{\sc i\kern-.025em b}\kern-.08em
    T\kern-.1667em\lower.7ex\hbox{E}\kern-.125emX}}
\begin{document}
\title{Impact of Source to Drain Tunneling on the Ballistic Performance of Ge, GaSb, and GeSn Nanowire p-MOSFETs }
\author{Dibakar Yadav, Deleep R. Nair,  \IEEEmembership{Member, IEEE}
\thanks{This paragraph of the first footnote will contain the date on 
which you submitted your paper for review.}
\thanks{D. Yadav and D. R. Nair are with Department of Electrical Engineering, IIT Madras, Chennai 600036, India  (e-mail: dibakaryadav19@gmail.com; deleep@ee.iitm.ac.in). }
 }

\maketitle

\begin{abstract}
We investigated the effect of material choice and orientation in limiting source to drain tunneling (SDT) in nanowire (NW) p-MOSFETs. Si, Ge, GaSb, and Ge$_{0.96}$Sn$_{0.04}$ nanowire MOSFETs (NWFETs) were simulated using rigorous ballistic quantum transport simulations. To properly account for the non-parabolicity and anisotropy of the valence band the k·p method was used. For each material, a set of six different transport/confinement directions were simulated to identify the direction with the highest ON-current ($I_{ON}$). For Ge, GaSb, and GeSn [001]/110/$\bar{1}$10 oriented NWFETs showed the best ON-state performance, compared to other orientations. Our simulation results show that, despite having a higher percentage of SDT in OFF-state than silicon, GaSb [001]/110/$\bar{1}$10  NWFET can outperform Si NWFETs. We further examined the role of doping in limiting SDT and demonstrated that the ON-state performance of Ge and GeSn NWFETs could be improved by reducing the doping in the source/drain (S/D) extension regions. Finally, we analyzed the impact of increased injection velocity in  [001]/110/$\bar{1}$10 oriented GaSb and GeSn NWFETs, as a result of the application of uniaxial compressive stress, and showed that when compared at a fixed OFF-current ($I_{OFF}$) with unstrained NWFETs, uniaxial compressive stress deteriorates the ON-state performance due to an increase in OFF-state SDT current component.
\end{abstract}


\section{Introduction}
\label{sec:introduction}
Increased SDT leakage in devices with short channel length can become a significant roadblock in scaling down transistor dimensions \cite{wang_iedm_02,meh_ted_13,jel_nt_15}. III-V semiconductors with high electron mobility like InGaAs, although regarded as promising candidates for future generation n-MOSFETs \cite{del_nat_11}, are more susceptible to SDT leakage due to their lower transport effective mass ($m^*_{trans}$). III-V channel based p-MOSFETs can be more immune to SDT leakage in OFF-state compared to their n-channel counterparts, at scaled gate lengths due to their lower hole mobility (higher $m^*_{trans}$) compared to electron mobility \cite{del_nat_11}. Devices based on III-V materials like GaSb are being actively explored as a potential candidate to replace Si as a channel for the future generation of p-MOSFETs \cite{dey_ecs_12,nai_ted_12}. At the same time, the anisotropic nature of the valence band makes the performance of scaled p-MOSFET devices strongly dependent on the direction of transport/confinement \cite{meh_wmed_09}. Hence it may be possible to engineer hole effective masses in materials with higher hole mobility compared to Si, to limit SDT and improve the device performance. Germanium used to have the highest bulk hole mobility among all the elemental group IV and  III-V semiconductors \cite{del_nat_11}. Recently, GeSn alloy based p-channel MOSFETs have achieved higher effective hole mobility compared to pure Ge based FETs \cite{han_iedm_11, gupta_iedm_11}. To enable device scaling with performance improvements over conventional Si-based p-MOSFETs, it is essential to explore the relative merits/demerits of MOSFETs based on alternate channel materials. Nanowire MOSFETs due to their ability to provide the ultimate electrostatic control of the channel by the gate are regarded as a promising device architecture to continue scaling \cite{kuhn_ted_12}. Hence in this paper, we have carried out a comparative analysis of the ballistic performance of Ge, GaSb, and GeSn NWFETs, to determine their suitability as a channel material for the future generation of p-MOSFETs.

A lot of studies have focussed in assessing the performance of  Si, Ge, and III-V nanowire n-MOSFETs in the presence of  SDT \cite{meh_ted_13,kim_ted_15,Lui_iedm_11,kim_iedm_15}.  But a similar study involving III-V materials along with Si, Ge for nanowire p-MOSFETs has not been carried out.  In \cite{syl_ted_12},  some  III-V materials alongside Si,  Ge NWs have been considered.  The authors have focussed on the ability of these materials in blocking SDT  current for n- and p-NWFETS,  but ON-state performance of these materials have not been evaluated.  Other studies involving nanowire p-MOSFETs have only considered Si and Ge as channel materials and have been carried out either at longer gate lengths \cite{kim_iedm_15}, with smaller SDT current component or have employed a semiclasical top of the barrier (ToB) model \cite{jing_iedm_05}, which doesn’t account for SDT.  In  \cite{neo_jap_10,meh_wmed_09}  ballistic performance of Si NWFETs has been evaluated in different transport orientations using the ToB model. In\cite{jel_nt_15}  an optimized range of m*  has been provided, which has been treated as a material independent quantity, to optimize device performance for  sub-12 nm nodes.  In\cite{jel_nt_15}   however, transport was treated using a single band effective mass (EM) model. The EM model can’t account for the non-parabolic and coupled nature of valence bands \cite{shin_jap_09}. Recently, Chang et al. \cite{chang_ted_17}  have analyzed the ballistic performance of  III-V double-gate p-MOSFETs using ToB semiclassical transport model.

In this work, we perform ballistic quantum transport simulations using the k·p method, to analyze the impact of SDT on the performance of nanowire p-MOSFETs. A comprehensive analysis of the effects of the valence band dispersion relations, resulting from the use of different channel materials and crystallographic orientations will provide essential guidelines in designing  sub-10 nm p-MOSFETs.  We have performed rigorous ballistic quantum transport simulations of NWFETs with Ge,  GaSb,  and  GeSn as the channel materials and compared their performance with Si  NWFETs.  For these materials, we have attempted to identify the transport directions which can minimize the  OFF-state  SDT without compromising too much on the ON-state performance.

The rest of the paper is organized as follows.  We briefly summarize the simulation approach in section II. In section III, we analyze the ballistic performance of all the four materials with different transport orientation,  strain, and source/drain doping concentrations.  We also investigate the behavior of injection velocity, quantum capacitance in the ballistic limit.  Finally, we conclude the paper in section IV.
  
\section{Approach}

To investigate the effect of SDT on the performance of different channel materials, we adopt the following methodology,
 \begin {enumerate}
\item  All the materials have been compared at a fixed gate length of $L_{G}=$10 nm. Dimensions of the NW cross-section are chosen to be 5 nm $\times$5 nm. Since our aim is to optimize $m^*_{trans}$ to limit SDT, by varying channel material and orientation, rather than by changing device geometry, the dimensions of the device cross-section were kept constant throughout all the simulations.   

\item We also analyzed the impact of channel transport orientation in minimizing SDT. For each material, NWFETs with six different transport orientations were simulated, to identify the orientation providing the highest $I_{ON}$. For a fair comparison, $I_{OFF}$ for each orientation was made 100 nA/$\mu$m, by adjusting the gate work function. 

\item The role of doping in minimizing SDT and improving $I_{ON}$ has been examined for Ge and GeSn NWFETs.

\item Finally, for the directions with highest $I_{ON}$, we examined the impact of compressive stress, to determine whether any further increase in the ballistic injection velocity translates into higher $I_{ON}$ for these devices.   
\end {enumerate}
\begin{figure}[t!]
\centerline{\includegraphics [width=0.8\linewidth, trim={0 5.3cm 0 5.5cm},clip] {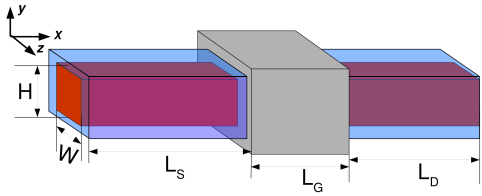}}
\caption{Schematic of the simulation domain of nanowire MOSFETs used in this work.}
\label{NW_fig}
\end{figure}
\begin{table}[h!]
	\centering
	  \caption{\textsc{Summary of NW axial orientation and directions of confinements }}
\begin{tabular}{@{}ccc@{}}
	\toprule
	\begin{tabular}[c]{@{}c@{}}NW axial\\  orientation\end{tabular} & \begin{tabular}[c]{@{}c@{}}Transport\\ direction (x)\end{tabular} & \begin{tabular}[c]{@{}c@{}}Confinement\\ directions (y/z)\end{tabular} \\ \midrule
	$[100]$                                                         & 100                                                               & 010/001                                                                \\
	$[110]$                                                         & 110                                                               & $\bar{1}$10/001                                                                \\
	$[111]$                                                         & 111                                                               & 0$\bar{1}1/\bar{2}$11                                                               \\
	$[001]$                                                         & 001                                                               & 110/$\bar{1}$10                                                                 \\
	$[0\bar{1}1]$                                                        & $0\bar{1}1$                                                            & $\bar{2}11$/111                                                               \\
	$[\bar{2}11]$                                                          & $\bar{2}11$                                                            & $0\bar{1}1$/111                                                               \\ \bottomrule
\end{tabular}
\label{ort_table} 
\end{table}
We have solved self-consistently, the 3D-Poisson's equation and Schr{\"o}dinger's equation within the non-equilibrium Green's function (NEGF) formalism, to analyze the effect of NW bandstructure and electrostatics on the overall performance of NWFETs. A schematic of NWFETs simulated in this study is shown in Fig.~\ref{NW_fig}. For all the materials considered in this study, transport characteristics with [100]/010/001, [110]/$\bar{1}$10/110, [111]/0$\bar{1}$1/$\bar{2}11$, [001]/110/$\bar{1}$10, [0$\bar{1}$1]/$\bar{2}11$/111, [$\bar{2}$11]/0$\bar{1}$1/111 orientations were simulated. The summary of NW transport directions simulated, with their directions of confinement is given in Table~\ref{ort_table}. Hereafter, for brevity, we denote different NWs using their direction of transport. For materials with indirect bandgaps, the 6 band $k\cdot p$ method provides an accurate description of valence bands around the $\Gamma$ point\cite{bal_JPD}. Hence for Si, Ge and GeSn, the 6 band $k\cdot p$ method has been used. For GaSb with a direct bandgap, we have used the 8 band $k\cdot p$ method. 

To reduce the computational load associated with the solution of NEGF equations, we first transformed the device Hamiltonian from real space to reciprocal space \cite{shin_jap_09}. Since the Hamiltonian in reciprocal space was still too expensive to be used in transport simulations, the mode-space (MS) Hamiltonian was constructed, which was then used in NEGF simulations. For the 6 band $k\cdot p$ method, the MS Hamiltonian was constructed following the approach outlined by Huang et al. in\cite{huang_ted_13}. Similar to \cite{huang_ted_13}, we constructed the MS Hamiltonian by sampling the modes at the $\Gamma$ point first (k-space sampling), and then by performing an energy space sampling at an energy of $E=E_{top}-E_{int}$, where $E_{top}$ is the energy at the top of the valence band edge and $E_{int}$ is the energy interval starting from $E_{top}$, over which we need the bandstructure obtained from the MS approach to match the bandstructure obtained using the reciprocal space Hamiltonian. The MS transformation Hamiltonian was then constructed by combining the modes obtained by k-space sampling to those obtained by energy space sampling and ortho-normalizing the resultant matrix\cite{huang_ted_13}. For GaSb with the 8 band $k\cdot p$ model, the approach proposed in \cite{huang_btbt_14} was used. For 8 band $k\cdot p$ method, only k-space sampling was used\cite{huang_btbt_14}. Spurious energy states in the MS Hamiltonian  were removed by discarding modes with singular values smaller than an iteratively determined threshold value \cite{huang_arxiv_15}. For simulating NWs with different transport/surface orientations appropriate coordinate transformations were
\begin{figure}[t!]
\centerline{\includegraphics [width=\linewidth, keepaspectratio,trim={0 0cm 0 0 cm},clip] {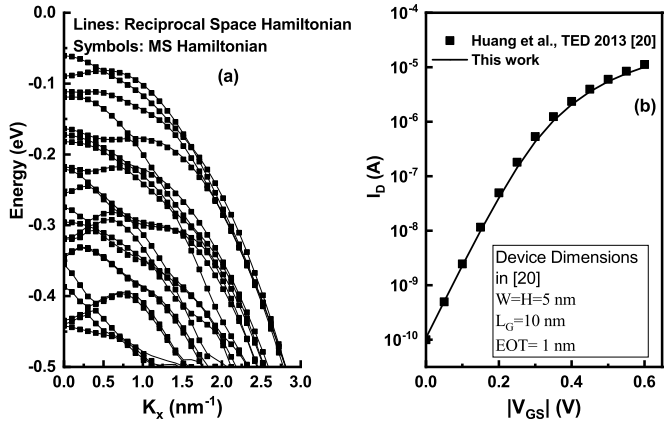}}
\caption{(a) Hole subbands of {Si} [100] nanowire with 5 nm$\times$ 5 nm area of cross-section,   obtained from  reciprocal space Hamiltonian (Solid lines) and mode space Hamiltonian (Symbols). (b) $I_{D}-V_{GS}$ characteristics for silicon NW pFET.  Benchmarked results of Si NWFET to compare  simulation approach. Our results (solid line) and \cite{huang_ted_13} (symbols)}
\label{ek_fig}
\end{figure}    
 performed\cite{esseni_palestri_selmi_2011}.  Recursive Green's function algorithm\cite{nik_06} was used to speed up the calculation of charge density. The converged charge density was then fed to a 3D Poisson's equation solver and these sequence of steps were repeated in a self-consistent manner.

 
 To check the validity of MS transformation, we have compared the E-k relation obtained by using the MS Hamiltonian to the one obtained using the reciprocal-space Hamiltonian. The bandstructure of [100] oriented Si NW is shown in Fig.~\ref{ek_fig}(a). To benchmark the NEGF simulation approach, we performed simulations of  a Si NWFET and compared it with a similar device in \cite{huang_ted_13}; the results of benchmarking are shown in Fig.~\ref{ek_fig}(b). The parameters used in simulation are given in Table~\ref{param_table}. Si parameters are taken from \cite{shin_ted_10}. Luttinger parameters for Ge and Sn are taken from \cite{gupta_ted_14}. Elastic stiffness constants of GeSn are linearly interpolated from parameters of Ge and Sn \cite{wang_ted_17}. Luttinger parameters of GaSb are taken from \cite{huang_arxiv_15} and elastic stiffness constants from \cite{chang_ted_17}.

\section{Results and Discussion}
The dimensions of the simulated devices are $L_{S}=L_{D}=15$ nm. $L_{G}= 10$ nm. The dimensions of the cross-section are $W=H=5$ nm. EOT of 0.6 nm and $V_{DD}=-0.5$ V were used in all simulations.  Doping levels in S/D extension regions are $10^{20}$ cm$^{-3}$ for Si, Ge, and GeSn NWFETs, and 5$\times$10$^{19}$ cm$^{-3}$ for GaSb NWFETs.
\subsection{Material Dependence}
\begin{figure}[t!]


\centerline{\includegraphics [width=\linewidth,trim={0 0cm 0 0 cm},clip] {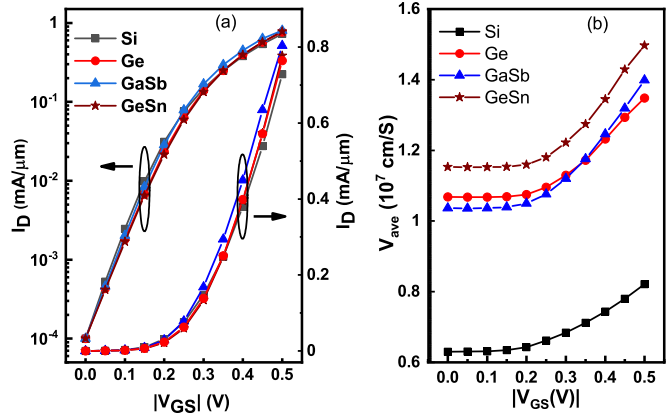}}

\caption{(a) $I_{D}-V_{GS}$ characteristics of Si, Ge, GaSb and GeSn [100] oriented NWFETs. (b) $V_{\!ave}$ at the virtual source for Si, Ge, GaSb and GeSn [100] oriented NWFETs .}
\label{cur_vel}
\end{figure}

\begin{figure}[t!]

\centerline{\includegraphics [width=\linewidth,trim={0 0cm 0 0 cm},clip] {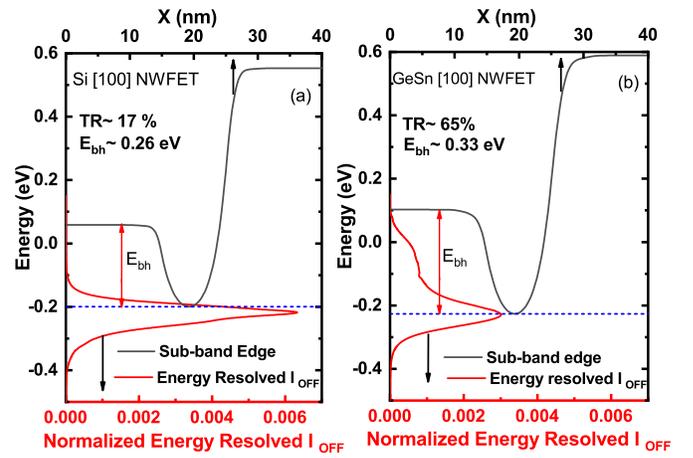}}

\caption{ Normalized current spectrum in OFF-state for  (a) Si and (b) GeSn [100] oriented NWFETs. Currents values have been normalized by total $I_{OFF}$. Current flowing above the blue dashed lines  constitutes the tunneling current.}
\label{cur_spec}
\end{figure}

In this subsection, we compare the ballistic performance of [100] oriented NWFETs for all the four materials. Figure~\ref{cur_vel}(a) shows the $I_{D}-V_{GS}$ characteristics of Si, Ge, GaSb and GeSn NWFETs oriented in [100] transport direction. Si NWFET has the lowest $I_{ON}$ due to its lower injection velocity. Figure~\ref{cur_vel}(b) shows the average ballistic injection velocity ($V_{\!ave}$) \cite{teherani_inj_17,zhang_cv_19} at the peak of the source-channel potential barrier. GeSn NWFET has the highest $V_{\!ave}$ among all the four materials. It also has the highest component of SDT in the OFF-state. Figure~\ref{cur_spec} shows the normalized energy resolved $I_{OFF}$ for Si and GeSn NWFETs in [100] orientation. For Si NWFET, tunnel ratio (TR) defined as the ratio of current flowing by tunneling to the total current in OFF-state is $\sim$17\%. Thus most of the current in OFF-state is due to thermionic emission over the potential barrier. GeSn NWFET on the other hand has a TR of $\sim$65\%, highest among all the four materials. Thus the potential barrier height ($E_{bh}$) needed to achieve the same $I_{OFF}$ in [100] orientted GeSn NWFET is higher compared to [100] Si NWFET. This results in lower ON-state overdrive in [001] orinted GeSn NWFET.

GaSb [100] NWFET has the highest $I_{ON}$ among all four materials. A lower TR$\sim$51\% for GaSb  NWFET compared to GeSn results in a higher ON-state overdrive.    
\begin{table}[t!]
\centering
\caption{\textsc{List of material parameters used in simulation. Only parameters needed during simulation are listed.}}
\begin{tabular}{@{}ccccccc@{}}
\toprule
Parameters & Si    & Ge    & GaSb  & GeSn   &  &  \\ \midrule
$\gamma _1$         & 3.60  & 9.37  & 13.27 & 10.54  &  &  \\
$\gamma _2$        & 0.67  & 3.01  & 4.97  & 3.38   &  &  \\
$\gamma _3$         & 1.21  & 4.02  & 5.978 & 4.52   &  &  \\
$\Delta_{\text{so}}$ (eV)   & 0.044 & 0.306 & 0.748 & 0.332  &  &  \\
$m_{c}$   & --      & --       & 0.042 &     --   &  &  \\
$E_{\text{g}}$ (eV)    & --      &  --     & 0.751 & --    &  &  \\
$E_{\text{p}}$(eV)    & --    & --    & 21.2  & --     &  &  \\
$a_{\text{c}}$ (eV)    & --      & --      & -7.5  & --       &  &  \\
$a_{\text{v}}$ (eV)    &       &       & 0.8   & 1.24   &  &  \\
b (eV)     & --      &  --     & -2.0  & -2.9   &  &  \\
d (eV)     & --      &   --    & -4.7  & -5.3   &  &  \\
C$_{11}$ (GPa)  &  --     &   --    & 88.5  & 123.72 &  &  \\
C$_{12}$ (GPa)  &   --    & --      & 40.2  & 43.41  &  &  \\
C$_{44}$ (GPa)   &    --   &   --    & 43.2  & 65.77     \\ \bottomrule                    
\end{tabular}
\label{param_table}
\end{table} GeSn NWFET has the worst SS initially, due to higher TR, implying degraded gate control over the channel. But once the devices start operating above the sub-threshold region, the rise in current as $V_{GS}$ increases, is steepest for GeSn NWFET. Reducing OFF-state tunneling by decreasing the doping concentration of S/D extension regions can counter the loss of ON-state overdrive. As shown in subsection~\ref{dop_sec} with similar S/D doping GeSn NWFET can outperform GaSb NWFET.

The performance of [100] oriented NWFETs is, however, sub-optimal in terms of $I_{ON}$. For all materials, an increase in the ballistic injection velocity, over its value in [100] oriented NWFETs results in an increased $I_{ON}$. However, for all materials $I_{ON}$ doesn't increase proportionately, with an increase in injection velocity. The orientation dependent performance variation for all the materials is discussed in the next subsection.

\begin{figure}[t!]
\centerline{\includegraphics [width=\linewidth,keepaspectratio,trim={0 0cm 0cm 0 cm},clip] {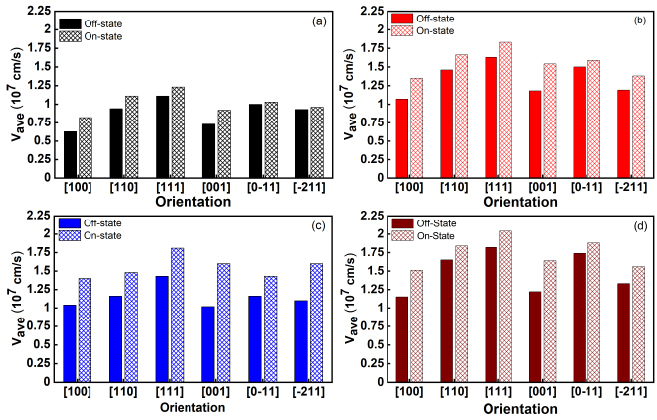}}
\caption{ON and OFF-state $V_{\!ave}$ at virtual source for (a) Si, (b) Ge, (c) GaSb, (d) GeSn NWFETs. }
\label{vel_bar}
\end{figure}
\subsection{Orientation Dependence}

\begin{figure}[t!]
\centerline{\includegraphics [width=\linewidth,keepaspectratio,trim={0 0cm 0cm 0 cm},clip] {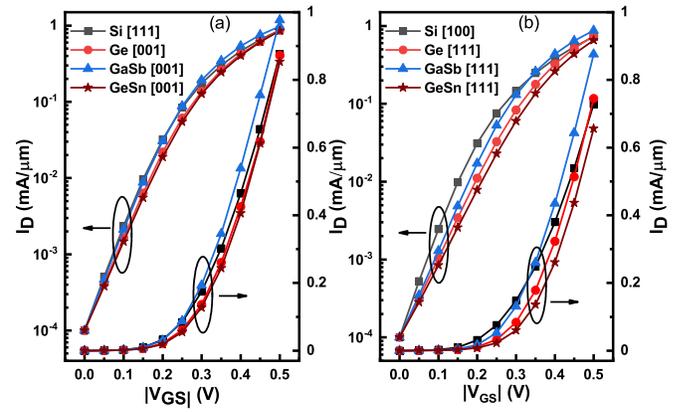}}
\caption{(a) $I_{D}-V_{GS}$ characteristics for orientations with highest $I_{ON}$ for each material. (b) Comparison of  $I_{D}-V_{GS}$ characteristics of [111] oriented Ge, GaSb, and GeSn NWFETs with [100] Si NWFET.}
\label{best_high}
\end{figure}

Table~\ref{I_on_ort} shows $I_{ON}$ for all materials with different orientations. Orientation dependence of  $V_{\!ave}$ at the virtual source is shown in Fig.~\ref{vel_bar}. Both ON and OFF-state ballistic average injection velocities are shown. Irrespective of the material choice, [111] oriented NWFETs have the highest $V_{\!ave}$, while [100] oriented NWFETs has the lowest $V_{\!ave}$. Figure~\ref{best_high}(a) shows the $I_D$-$V_{GS}$ characteristics of orientations with the highest ON-current for each material. In the case of Si NWFETs, [111] oriented NWFET has the highest $I_{ON}$. The superior performance of Si [111] oriented NWFET, which has the highest   $V_{\!ave}$ and TR among all Si NWFETs, shows that all Si NWFETs still operate in the thermionic current component dominated regime in OFF-state. 
\begin{table}[ht!]
	\centering
	 \caption{$\textit{I}_{ON}$ \textsc{for} {Si, Ge, GaSb}, \textsc{and} {GeSn} {NWFETs} \textsc{with different transport orientation}}

	\begin{tabular}{@{}ccccc@{}}
		 \toprule
		\multirow{2}{*}{Orientation} & \multicolumn{4}{c}{$I_{ON}$ (mA/um)} \\
		\cmidrule(l){2-5} 
		& Si     & Ge     & GaSb  & GeSn  \\
		\cmidrule(l){1-5}
		100/010/001                  & 0.73   & 0.76   & 0.80  & 0.78  \\
		110/$\bar{1}$10/001                 & 0.84   & 0.74   & 0.80  & 0.69  \\
		111/0$\bar{1}$1/$\bar{2}$11                & 0.88   & 0.75   & 0.88  & 0.66  \\
		001/110/$\bar{1}$10                 & 0.80   & 0.87   & 0.98  & 0.85  \\
		0$\bar{1}$1/$\bar{2}$11/111                & 0.80   & 0.70   & 0.82  & 0.62  \\
		$\bar{2}$11/0$\bar{1}$1/111                & 0.82   & 0.72   & 0.85  & 0.62 \\ \bottomrule
		
	\end{tabular}
\label{I_on_ort}
\end{table}
For Ge, GaSb, and GeSn [001] oriented NWFETs have  higher $I_{ON}$, compared to other orientations. The reason for the superior performance of [001] oriented NWFETs, over [100] oriented NWFETs is their very similar $V_{\!ave}$ in OFF-state to [100] oriented NWFETs, as shown in Fig.~\ref{vel_bar}. This results in [001] oriented NWFETs having similar TR and SS to [100] oriented NWFETs.  In ON-state, however, the injection velocity for [001] oriented NWFETs is much higher compared to [100] orientation, thus resulting in better ON-state performance. In ON-state, as k-states with higher energy, farther away from the Brillouin zone center are occupied, $V_{\!ave}$ increases. These states have higher velocity compared to states near the $k=0$ point \cite{teherani_inj_17}, due to the higher gradient of the dispersion relation for off-zone center states. The gradient of energy with respect to k for [001] oriented NWs for Ge, GaSb, and GeSn is much higher compared to [100] oriented NWFETs. Hence once the high energy off-zone centre states are populated in ON-state, $V_{\!ave}$. increases significantly for [001] NWFETs.

  For [111]    oriented NWFETs, the larger injection velocity  (lower $m^*_{trans}$) \begin{figure}[t!]

\centerline{\includegraphics [width=\linewidth] {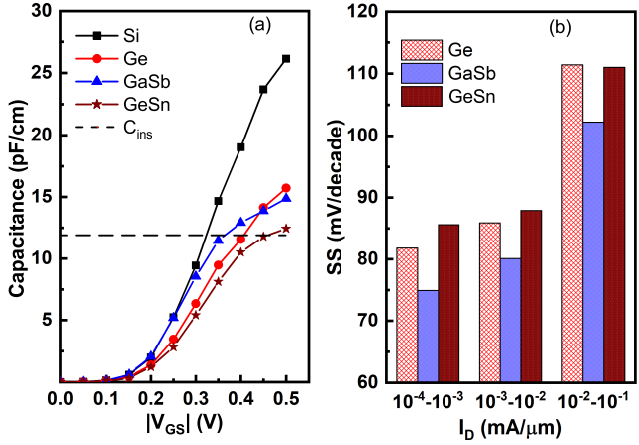}}

\caption{(a) Quantum capacitance vs Gate voltage for directions with highest ON-current for each material. (b) SS for first 3 decades of I$_D$  for [001] oriented   Ge,GaSb and GeSn NWFETs.}
\label{cap_and_ss}
\end{figure}  
  \begin{figure}[t!]

\centerline{\includegraphics [width=\linewidth,trim={0 0cm 0 0 cm},clip] {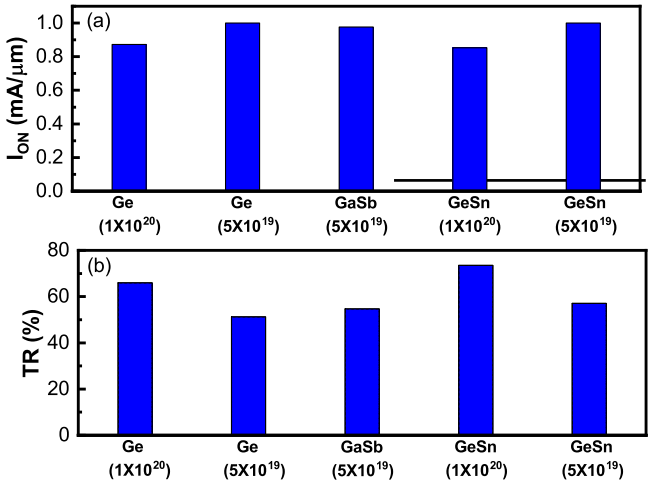}}

\caption{(a) Doping dependence of $I_{ON}$ for Ge and GeSn [001] oriented NWFETs. $I_{ON}$ for [001] oriented GaSb NWFET is also shown for comparison.  (b) Tunnel ratios for devices considered in (a). Doping levels in S/D extension regions in cm$^{-3}$ are indicated within brackets.}
\label{on_tunnel}
\end{figure} results in higher SDT, and degraded gate control. Figure~\ref{best_high}(b) shows the $I_D-V_{GS}$   characteristics of [111] oriented NWFETs, for Ge, GaSb and GeSn NWs. $I_D-V_{GS}$ characteristic of Si [100] NWFET, which has the lowest $V_{\!ave}$ in ON and OFF-states is also shown for comparison. The degraded gate control and higher SS for [111] NWs results in them  showing much inferior performance in the sub-threshold region. The SS for the first decade of change in $I_{D}$ from the $I_{OFF}$ value, and TR for different materials and transport direction combinations, is given in Table ~\ref{ss_tr_table}. As can be seen, for all the materials [111] oriented NWFETs have the worst SS and highest TR.  Higher  $V_{\!ave}$ for [111] oriented NWFETs is not enough to compensate for the loss of ON-state overdrive due to higher SDT in OFF-state. Hence [111] oriented Ge, GaSb, and GeSn NWFETs underperform compared to [001] oriented NWFETs for these materials.

For comparison across materials, we show the quantum capacitance (QC)\cite{kha_ted_09} as a function of $V_{GS}$ in Fig.~\ref{cap_and_ss}(a) for Si [111] and [001] oriented Ge, GaSb, and GeSn NWFETs. These are the directions having the highest $I_{ON}$ for each material. Our simulation results show that these devices operate in the classical capacitance regime\cite{kha_ted_09}. Hence the QC is greater than the NW insulator capacitance ($C_{ins}$)\cite{luna_sse_08}. Ge and GaSb [001] oriented NWFETs have comparable values of QC and $V_{\!ave}$ (as shown in Fig.~\ref{vel_bar}). Despite this, GaSb [001] NWFET outperforms Ge [001] oriented NWFET and has the highest $I_{ON}$ among all materials, with all different orientations considered.  Ge [001] oriented NWFETs underperform compared to GaSb [001] oriented NWFETs primarily due to higher OFF-state SDT and worse SS. Figure~\ref{cap_and_ss}(b) shows the SS over the first three decades of $I_D$, over which the characteristics are sub-threshold like. Ge [001] NWFET has higher SS in this region due to degraded gate control as a result of higher SDT. Higher SDT in Ge [001] NWFET compared to GaSb [001] oriented NWFET is partly also due to the higher doping in the S/D extension regions. SDT can be reduced by reducing the doping in S/D extension regions \cite{kim_eds}. With the same level of doping as GaSb, both Ge and GeSn NWFETs outperform [001] GaSb NWFETs, as shown in the next subsection.
\begin{table}[t!]
	\centering
	\caption{\textsc{SS and TR for} {Si, Ge, GaSb,} \textsc{and} {GeSn NWFETs} \textsc{with different transport orientations}}
	\begin{tabular}{@{}cccccc@{}}
		\toprule
\begin{tabular}[c]{@{}c@{}}Material\\ and Orientation\end{tabular} & \begin{tabular}[c]{@{}c@{}}SS\\ (mV/dec)\end{tabular} & TR      &  &  &  \\ \midrule
		Si $[100]$                                                         & 70                                                    & 16.64\%    &  &  &  \\
		Si $[111]$                                                         & 72                                                    & 39.76\%  &  &  &  \\
		Ge $[001]$                                                         & 82                                                    & 65.97\% &  &  &  \\
		Ge $[111]$                                                         & 99                                                    & 91.24\% &  &  &  \\
		GaSb $[001]$                                                       & 75                                                    & 54.69\% &  &  &  \\
		GaSb $[111]$                                                       & 90                                                    & 86.23\% &  &  &  \\
		GeSn $[001]$                                                       & 86                                                    & 73.39\% &  &  &  \\
		GeSn $[111]$                                                       & 108                                                   & 95.86\% &  &  &  \\ \bottomrule
	\end{tabular}
\label{ss_tr_table}
\end{table} 
\subsection{Impact of Doping and Strain} \label{dop_sec}
\begin{figure}[t!]

\centerline{\includegraphics [width=\linewidth,trim={0 0cm 0 0 cm},clip] {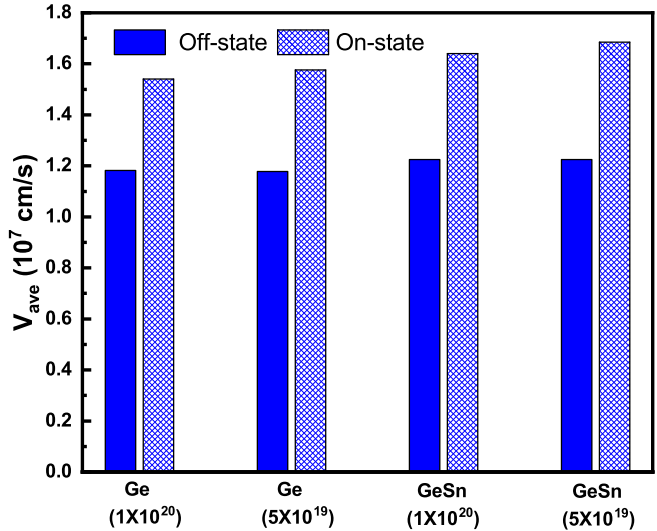}}

\caption{(a) Doping dependence of  ON and OFF-state $V_{\!ave}$ at the virtual source for [001] oriented Ge and GeSn NWFETs having two different doping levels in S/D extension regions.}
\label{low_vel}
\end{figure}

In this subsection, we reduce the doping levels in the S/D extension regions of Ge and GeSn NWFETs from $1\times 10^{20}$ to $5\times 10^{19}$ $\mathrm{cm^{-3}}$, to compare their performance with GaSb NWFETs at the same level of doping. [001] oriented NWFETs were simulated as they provide the highest $I_{ON}$ for each of the  three materials. Figure~\ref{on_tunnel}(a) shows $I_{ON}$ for [001] oriented Ge and GeSn NWFETs with S/D doping of 5$\times 10^{19}$ $\mathrm{cm^{-3}}$ and 1$\times 10^{20}$ $\mathrm{cm^{-3}}$. At the same value of S/D doping, Ge and GeSn NWFETs show marginally better ON-state performance as compared to GaSb [001] oriented  NWFET.   The TRs for these devices is shown in Fig.~\ref{on_tunnel}(b). Tunnel ratios improve considerably for both Ge and GeSn [001] oriented NWFETs at lower doping levels. Reduction of S/D doping leads to lower SDT and improved SS. The impact of lower S/D doping on $V_{\!ave}$ at the virtual source is shown in Fig.~\ref{low_vel}. The OFF-state $V_{\!ave}$ remains practically unchanged for lower doping levels in the S/D extension regions. Hence, the reduction in SDT is due to the widening of source-channel potential as a result of lower doping levels in S/D extension regions\cite{jel_nano_15}. The ON-state $V_{\!ave}$ increases slightly, primarily due to enhanced ON-state overdrive voltage, which results in hole sub-bands moving more closer to the  source contact Fermi level, thus increasing the occupation probability of the sub-bands.

 Finally, we have also examined the impact of compressive stress on the performance of GaSb and GeSn [001] oriented NWFETs, to determine whether any further increase  in the injection velocity due to strain, benefits the ON-state performance.  \begin{figure}[t!]

\centerline{\includegraphics [width=\linewidth,trim={0 0cm 0 0 cm},clip] {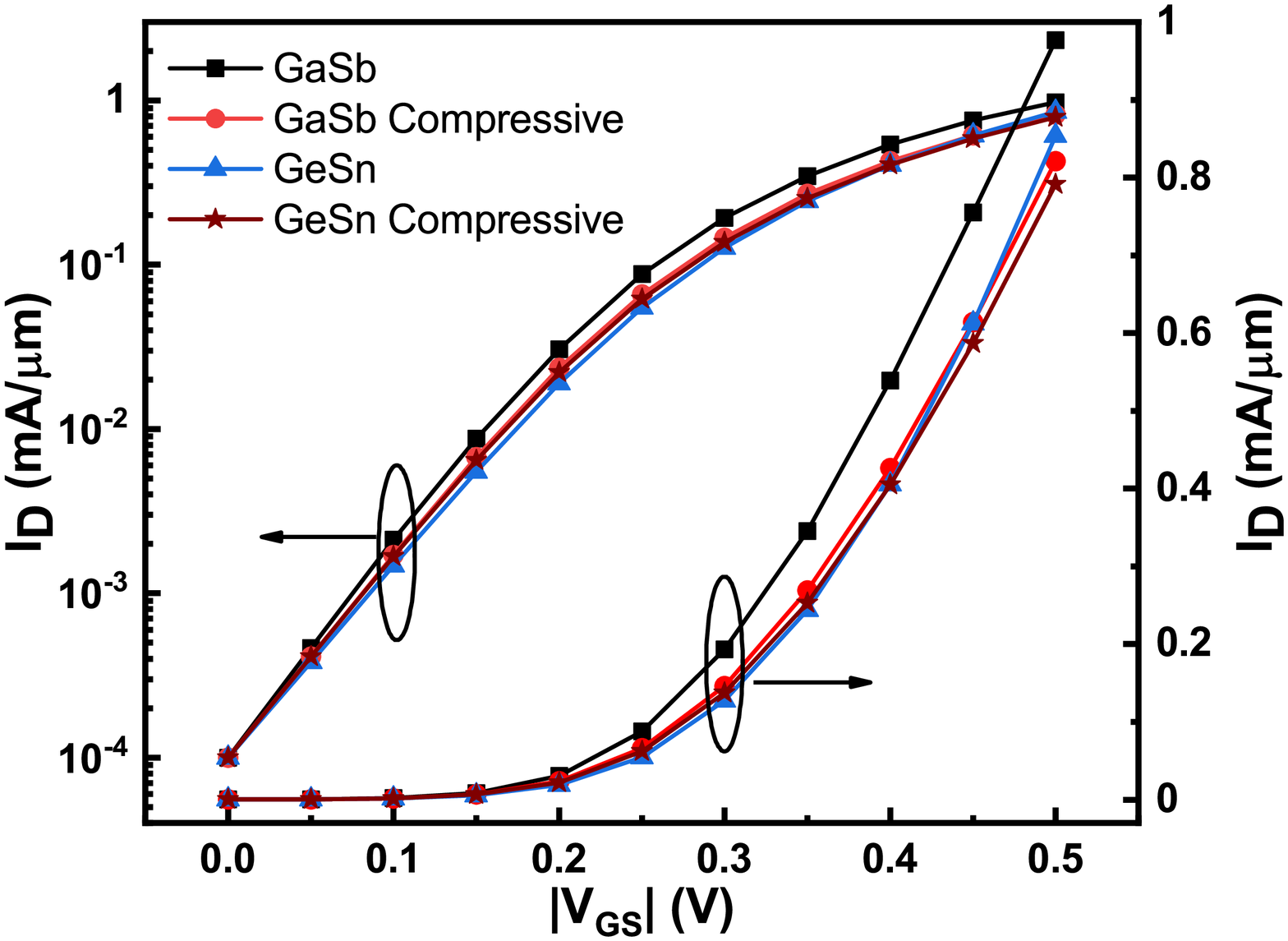}}

\caption{(a) I$\mathrm{_{D}}$-V$\mathrm{_{GS}}$ characteristics of  [001]  oriented GaSb and GeSn NWFETs with and without uniaxial compressive stress of 1 GPa.}
\label{comp_id}
\end{figure}  \begin{figure}[ht!]


\centerline{\includegraphics [width=\linewidth ] {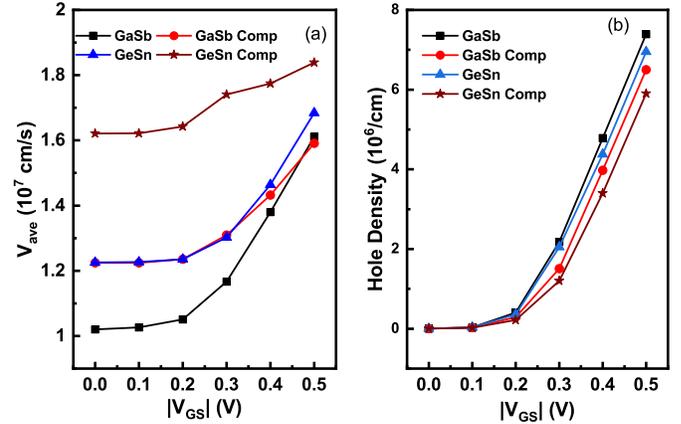}}

\caption{(a) $v_{ave}$ at the virtual source of  [001] oriented GaSb and GeSn NWFETs with and without uniaxial compressive stress. (b) Inversion charge density at the virtual source for devices considered in (a).}
\label{comp_inv}
\end{figure} Ge [001] NWFET has not been considered as it has similar value of  $I_{ON}$, elastic constants, and deformation potentials as GeSn [001] NWFET.  Our simulation results show that ON-state performance deteriorates due to uniaxial compressive stress. Figure~\ref{comp_id} shows the $I_{D}-V_{GS}$ characteristics of GaSb and GeSn [001] NWFETs with 1 GPa compressive stress along transport direction, the   $I_{D}-V_{GS}$ characteristics of GaSb and GeSn [001] unstrained NWFETs is also shown for comparison. OFF-state SDT is enhanced due to uniaxial compressive stress, resulting in inferior ON-state performance. Unlike the case of reduced doping, the OFF-state  $V_{\!ave}$ is increased due to compressive stress, which results in higher SDT in OFF-state. Average ballistic injection velocity as a function of $V_{GS}$ is shown in Fig.~\ref{comp_inv}(a) for both GeSn and GaSb NWFETs with and without applied compressive stress. As can be seen, the OFF-state $V\mathrm{_{ave}}$ increases significantly for compressively strained GeSn and GaSb NWFETs, resulting in increased SDT. Figure~\ref{comp_inv}(b) shows the inversion charge density at the virtual source for strained and unstrained GaSb and GeSn NWFETs. The inversion density is also slightly reduced due to compressive stress. Thus any further injection velocity enhancement doesn't result in better ON-state performance at this gate length, when compared at a fixed $I_{OFF}$.
\section{Conclusion}
In summary, we have studied the effectiveness of material choice and orientation as handles to counter SDT in OFF-state. Our simulation results show that,

\begin{enumerate}
\item At $L_G=10$ nm, all Si NWFETs have SDT current component $< 50\%$ in OFF-state. Among all Si NWFETs, [111] oriented NWFET with highest $V_{\!ave}$ shows the best  ON-state performance.
\item Other materials operate in the tunneling current dominated regime in OFF-state $(TR>50\%)$. For Ge, GaSb and GeSn, [001] oriented NWFETs with $V_{\!ave}$ lying between that of [100] and [111] orientation for each material, show the best ON-state performance.
\item GaSb [001] oriented NWFET shows the best ON-state performance, among all the materials with all six different transport orientations considered. This is true, when Ge and GeSn NWFETs have higher doping concentration in S/D extension regions. At the same level of doping; however, [001] oriented Ge and GeSn 
NWFETs perform slightly better as compared to GaSb NWFET with the same orientation. 

\item Finally, our simulation results show that any further enhancement in the injection velocity for [001] oriented NWFETs with GaSb and GeSn as the channel materials, by the application of uniaxial compressive stress doesn’t result in better ON-state performance when compared at a fixed $I_{OFF}$. In fact, due to increased SDT in OFF-state, $I_{ON}$ is reduced in [001] oriented NWFETs with uniaxial compressive stress.

 \end{enumerate}
We note here that the impact of the NW cross-section dimensions, on the device performance,  has not been considered in this work. For NWs with smaller cross-section, atomistic tight binding (TB) method will be more accurate compared to the continuum based $k\cdot p$ method. For them, the modes-space transformation of the TB Hamiltonian \cite{mil_prb_12} can also be performed at a relatively cheaper computational cost. The impact of NW cross-section dimension and gate length on these set of channel materials will be the subject of a future study. While in this work, we have not considered the effects of phonon and surface roughness scattering and alloy scattering for GeSn NWFETs, the current study can nonetheless provide useful guidelines in the selection of materials and orientations, so as to minimize the effect of SDT in nanowire p-MOSFETs.
\bibliographystyle{IEEEtran}

\end{document}